# Amorphous Boron Nitride: *Ab initio* Study of its vibrational properties

David Hinojosa-Romero[1], Alexander Valladares[2], Renela M. Valladares[2], Isaías Rodríguez[1], and Ariel A. Valladares[1] *

[1] Instituto de Investigaciones en Materiales, Universidad Nacional Autónoma de México, Apartado Postal 70-360, Ciudad Universitaria, CDMX, 04510, México.

[2] Facultad de Ciencias, Universidad Nacional Autónoma de México, Apartado Postal 70-542, Ciudad Universitaria, CDMX, 04510, México

*Corresponding Author: Ariel A. Valladares, e-mail, valladar@unam.mx

**Abstract**

**Boron nitride (BN) is a structurally versatile insulator since it can be found in several crystalline structures with interesting mechanical and electrical properties, making this material attractive for technological applications. Seeking to improve its features, experimental and simulational studies for the amorphous phase ($a$-BN) have been carried out by some groups, focusing on the electrical and structural properties, pressure-induced phase transformations, and hydrogenated $a$-BN. In this work two amorphous structures are computationally generated and studied using *ab initio* Molecular Dynamics on a 216-atom supercell with two different densities, 2.04 and 2.80 g cm$^{-3}$. Our *undermelt-quench* approach is followed, since it has proven to give good structures for disordered materials and their properties. The topology, the vibrational density of states and some thermodynamic properties of the two samples are reported and compared with existing experiment. Some computational results are also revisited.**

## DECLARATIONS

**Funding** Financial support was received from DGAPA-UNAM (PAPIIT) under projects No. IN104617 and IN116520.

**Conflicts of interest/Competing interests.** The authors declare they have no competing interests to declare that are relevant to the content of this article.

**Availability of data and material.** The datasets generated and analyzed during the current study are available from the corresponding author on reasonable request.

**Code availability**. Code Correlation can be found in the GitHub repository: https://github.com/Isurwars/Correlation (DOI: 10.5281/zenodo.5514113).

**Authors' contributions** Ariel A. Valladares, Alexander Valladares and Renela M. Valladares conceived this research and designed it with the participation of David Hinojosa-Romero and Isaías Rodríguez. All the simulations were done by David Hinojosa-Romero and Isaías Rodríguez. All authors discussed and analyzed the results. Ariel A. Valladares wrote the first draft and the other authors enriched the manuscript.

# INTRODUCTION

Boron is an atom with the electronic structure dictated by a He-like core, plus 2 electrons in the *2s* shell and 1 electron in the *2p* shell; atomic nitrogen, located on the same row of the Periodic Table, has also a He-like core plus 2 *2s* and 3 *2p* electrons. When alloyed in equal quantities in the solid state, BN, 4 crystalline structures may appear, two three-dimensional (3-d) and two two-dimensional (2-d). In the bulk, one has 3-d forms like the cubic BN (*c*-BN), which has the zinc blende structure and another structure, rare hexagonal, the wurtzite (*w*-BN). *c*-BN is a very hard material since it is tetrahedral and covalently bonded; *w*-BN is also covalently bonded in a hexagonal arrangement. In 2-d one has the layered hexagonal BN (*h*-BN) and the rhombohedral BN (*r*-BN) structures. The *h*-BN structure is the most stable and consists of alternating B and N atoms in layers, covalently bonded and its name is due to the hexagonal rings formed by the elements. *r*-BN consists of three layers arranged in a stacking structure. In both cases the van der Waals forces are responsible for maintaining the layers together [1].

However, boron nitride in the amorphous phase (*a*-BN), may lose all characteristics specific to the several crystalline structures, 3-d or 2-d, although the density of these structures may influence the type of arrangements in the amorphous and also the hybridization of the bonds. Therefore, the density of *a*-BN may be relevant to expect some manifestation of the crystalline forms. The amorphous phase of BN has been studied less than the crystalline topologies to the point that specific basic properties like the correlation functions are scarce. Also, although the thermal properties due to the disordered lattices are important to understand the behavior of the material, as far as the authors know not much has been done along these lines. One of the reasons for this deficit in knowledge may be due to the difficulty to study, both experimentally and simulationally, the amorphous phase. The first attempt recorded is due to Rand and Roberts [2] and Hirayama and Shohno [3] who produced samples of "clear, vitreous films" [4] reminiscent of disordered materials. Papers have appeared along the years and one could mentioned some of the most recent ones that make reference to previous works [5].

Concerning the vibrational density of states, even less has been done. Recently, Wu and Han performed a calculation of the thermal conductivity of a defective-to-amorphous hexagonal boron nitride monolayer, and they obtained some vDoS customizing the nature of the defects [6]. The classical empirical Tersoff potential was used as an approximation to the quantum nature of the interatomic interactions. Vibrational densities of states were obtained for some specific two-dimensional structures and their relevance to the thermal conductivity was studied.

After the limited number of related works, it was decided that our approach should be different from the ones reported in the current literature. The density was considered the relevant factor and based on it, 3-d atomic topologies were generated for two values of the density. The initial structure was an unstable one as demanded by our *undermelt-quench* approach [7], a procedure that has generated amorphous samples congruent with reality [8], which implies that no crystalline structure of BN for the selected densities was the starter. The specimens were heated to just below the liquidus to avoid liquid-like arrangements in the final disordered specimens, despite the similarity between these two phases. The densities chosen were 2.04 and 2.80 g cm$^{-3}$ in accordance with the experimental values of 1.7 to 2.1 g cm$^{-3}$ commonly used [4]. Also, the pioneering simulations of Mc Culloch *et al.* considered two densities as well, 2.0 and 3.0 cm$^{-3}$ [9] and was the first application of Car-Parrinello density functional theory to generate the amorphous phase of BN by cooling down from the melt a supercell with 64 atoms.

With this in mind we proceeded to the creation of the amorphous samples and studied them by obtaining the Pair Distribution Function (PDFs), the reduced PDFs (rPDFs), the partial PDFs (pPDFs), the Plane Angle Distribution Functions (PADs), the Vibrational Density of States (vDoS, or *F(ω)*), and some thermodynamic properties due to the vibrations, such as the internal energy and the specific heat at constant volume, as a function of temperature. This is what we report and analyze in this work.

**METHOD**

The amorphous samples were generated by means of *ab initio* Molecular Dynamics (MD), and the *undermelt-quench* approach, a procedure developed in our group [7,8]. This approach considers a starting unstable supercell for the system which helps to the disordering process of the MD. The unstable structure is subjected to a heating ramp which takes the system, in 100 steps, from 300 K to a temperature below the liquidus, followed by a cooling ramp that reaches 0 K as close as possible; the maximum temperature is below the liquidus in order to prevent final structures with liquid characteristics. The heating/cooling slopes are, in absolute value, the same. After the MD, a Geometry Optimization (GO) process is carried out to allow the atoms accommodate in their local minimum energy positions. The MD and GO processes were done for the two supercells reported in this work. The calculated correlation functions were produced with **Correlation**, a code developed by Rodríguez *et al.* [10] and were compared with experiments and simulations.

Once the amorphous structures were obtained, vibrational frequencies for each system were calculated by the frozen-phonon method which consists in evaluating the mass-weighted Hessian (second energy derivatives) matrix and its subsequent diagonalization in order to obtain the vibrational eigenmodes for the system [11]. The vDoS corresponds to the number of eigenmodes within an energy interval (*E, E + δE*). The *F(ω)*s reported in this paper are normalized to 1 and, as functions of energy (*E = ℏω*), have units of states per meV.

Since the contribution of the lattice vibrations to the thermodynamic properties of materials as a function of temperature can be estimated through the calculated vDoS [12,13], the internal energy, *ΔE*, and the constant-volume specific heat, $C_v$, are also reported. The evaluations of both thermodynamic functions were done following the equations [14]:

$$\Delta E = 3nN\frac{\hbar}{2}\int_0^{\omega_L} \omega \coth\left(\frac{\hbar\omega}{2k_BT}\right) F(\omega)\,d\omega, \qquad (1)$$

$$C_v = 3nNk_B \int_0^{\omega_L} \left(\frac{\hbar\omega}{2k_BT}\right)^2 \operatorname{csch}^2\left(\frac{\hbar\omega}{2k_BT}\right) F(\omega)\,d\omega, \qquad (2)$$

where *n* is the number of atoms per unit cell, *N* is the number of unit cells, $\omega_L$ is the largest phonon frequency, and $F(\omega)$ is the normalized vDoS.

## CALCULATION SPECIFICS

All the simulations were spin-unrestricted all-electron calculations done within the Density Functional Theory framework as implemented in the DMol[3] code [15] included in the Materials Studio software [16]. Atomic orbitals for boron and nitrogen were described by a Double Numerical plus polarization d-function basis (DND) with an orbital cutoff of 4.1 Å. The exchange-correlation functional was treated under the Local Density Approximation (LDA) parameterized by Vosko, Wilk and Nusair (VWN) [17]. The Self-Consistent-Field density convergence threshold was set to 1 x 10$^{-6}$ and all the numerical integrations were done with a fine integration grid (as defined by DMol[3] [15]).

For the MD processes, the starting structures were two distinct unstable diamond-like substitutionally disordered supercells containing 108 boron atoms and 108 nitrogen atoms, resulting in a 216-atom cubic supercell of edge length of 12.96 Å for the supercell with a density of 2.04 g cm$^{-3}$, and 11.67 Å for the density of 2.80 g cm$^{-3}$ [4]. The initial velocities assigned to each atom followed a Maxwell-Boltzmann distribution at 300 K; the subsequent temperature control for the MD was realized within an NVT ensemble by means of the Nosé-Hoover thermostat [18,19] with a 0.5 Q-ratio parameter. The *undermelt-quench* process started from 300 K and reached, in 100 steps, a maximum temperature of 2000 K, since the liquidus temperature is 2380 K for the *h-BN* system. The cooling ramp followed immediately and reached 5.5 K in 118 steps. The time step for the MD process was 1.5 fs, resulting in a total simulation time of 327 fs. The GO processes were carried out using delocalized internal coordinates with the following convergence thresholds: 2.72 x 10$^{-4}$ eV for energy, 5.44 x 10$^{-2}$ eV Å$^{-1}$ for maximum force, and 5 x 10$^{-3}$ Å for maximum displacement.

For the calculation of the vibrational eigenmodes, the mass-weighted Hessian matrices were computed numerically using finite differences of first energy derivatives with a step size of 0.005 Å. The vDoS were found smoothing with a 3-point Fast-Fourier Transform (FFT) the phonon energy distributions, which were obtained from counting the number of eigenmodes of each supercell located within a 1.5 meV interval.

## RESULTS AND DISCUSSION

Amorphizing covalently-bonded materials is a challenge. In particular, solids like carbon reach different final structures and different bonding depending on the density of the starting material [20]. Also, the fact that the melting temperatures of this covalent solids are in the thousands of kelvins, and the fact that some methods liquefy the specimens first and then quench it from the liquid, involves the consideration of very high temperatures and the possible appearance of liquid characteristics in the final structures. The approach could be quantum mechanical or classical [21] and even though with classical potentials it is possible to deal with thousands of atoms in a supercell, the description is not as adequate as the quantum mechanical one that better represents the nature of the chemical bond.

In our *undermelt-quench* approach [7,8], an unstable periodic structure is chosen at the start arranging the atoms so as to give the experimental density that is investigated. No melting of the material is involved and the quantum methods are used to better describe the electronic structure. Our calculations, described in the previous section, will be presented now.

Figures 1 represent the MD and GO processes. Figure 1(a) depicts the heating and cooling disordering process, whereas Figure 1(b) describes the GO process where the total energy of the system diminishes until a locally stable structure is reached. As a direct consequence of the energy minimization of the GO processes, the binding energy ($E_b$) increments up to 6.784 and 6.776 eV per atom for the 2.04 and 2.80 g cm$^{-3}$ densities, respectively; those values should be taken with caution for the known overestimation of binding energies in the LDA [13].

The resulting structures after the GO processes are portrayed in the insets of Figure 1(b), where boron and nitrogen are represented by spheres of different colors, light for boron and dark for nitrogen. By looking at the distribution of the atoms within the supercell one can appreciate the stochastic distribution of the constituents.

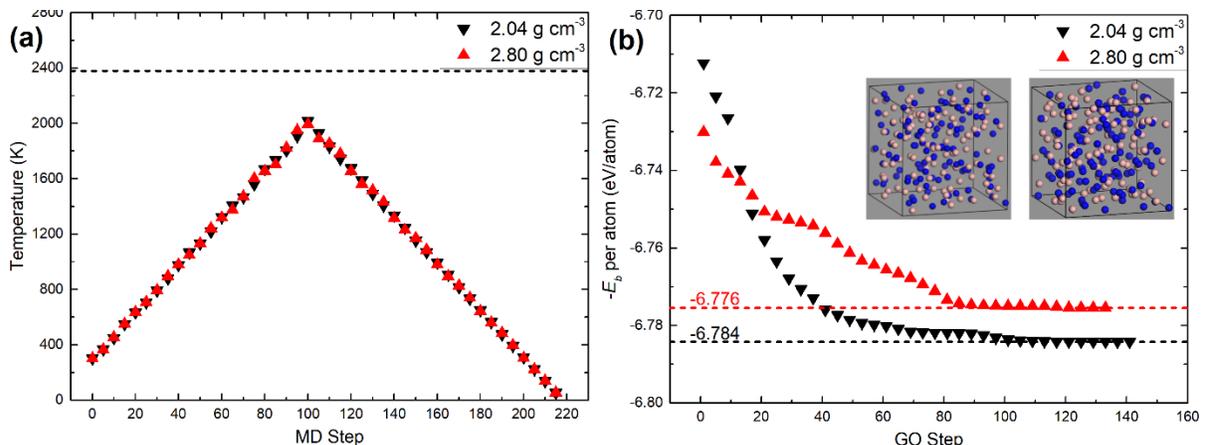

Figure 1. Generation of the amorphous structures. (a) The MD processes according to the *undermelt-quench* approach. The broken horizontal line corresponds to the liquidus temperature. (b) The GO processes in which -$E_b$ is plotted against the GO step. The resulting supercells are also shown (left for 2.04 g cm$^{-3}$, right for 2.80 g cm$^{-3}$), boron is represented by the light spheres whereas nitrogen are the dark spheres.

### i) Correlation functions

In Figures 2 the PDFs, total and partial, for the two samples are depicted. As shown in Figure 2(a) and Table 1, the positions for the first two peaks of the total PDF are the same regardless of the density, which implies that in the amorphization processes the interatomic distances tend to be the same; a decrease in density would probably generate pores within the sample without changing much the interatomic distances. The partial PDFs, which are shown in Figures 2 (b), (c), and (d), also manifest this tendency; however, the differences in the positions for the first peaks of N-N and B-B (see Table 1) may have an influence in the high-frequency modes of the samples.

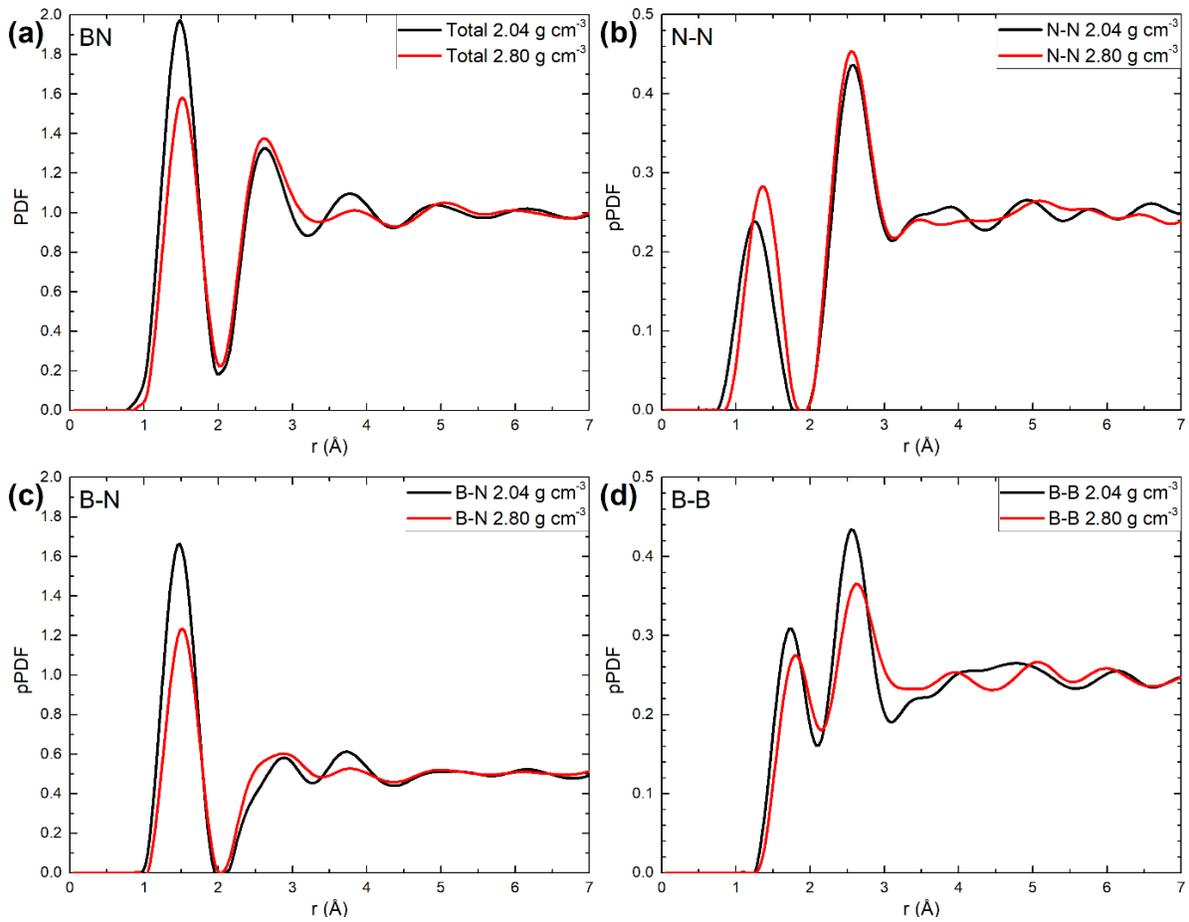

Figure 2. (a) Total and weighted partials b) N-N, c) B-N, and d) B-B PDFs for both densities. Black lines correspond to the density of 2.04 g cm$^{-3}$; red lines, to the density of 2.80 g cm$^{-3}$.

Table 1. Positions for the first two peaks of the PDFs and pPDF for both samples.

| Peak | | 1st peak (Å) | | 2nd peak (Å) | |
|---|---|---|---|---|---|
| Density | | 2.04 g cm$^{-3}$ | 2.80 g cm$^{-3}$ | 2.04 g cm$^{-3}$ | 2.80 g cm$^{-3}$ |
| PDF | | 1.5 | 1.5 | 2.6 | 2.6 |
| pPDF | N-N | 1.3 | 1.4 | 2.6 | 2.6 |
| | B-N | 1.5 | 1.5 | 2.9 | 2.9 |
| | B-B | 1.7 | 1.8 | 2.6 | 2.6 |

Figure 3 represents a comparison of our rPDFs, G(r), and those reported by experimentalists [5,22]. The discrepancies are notable, especially with the experimental results of Hong, a very recent work. It is surprising, taking into account the potential technological uses of *a*-BN, that there are such few correlation functions determined for the amorphous. One would expect that more calculations should have been carried out for these amorphous phases.

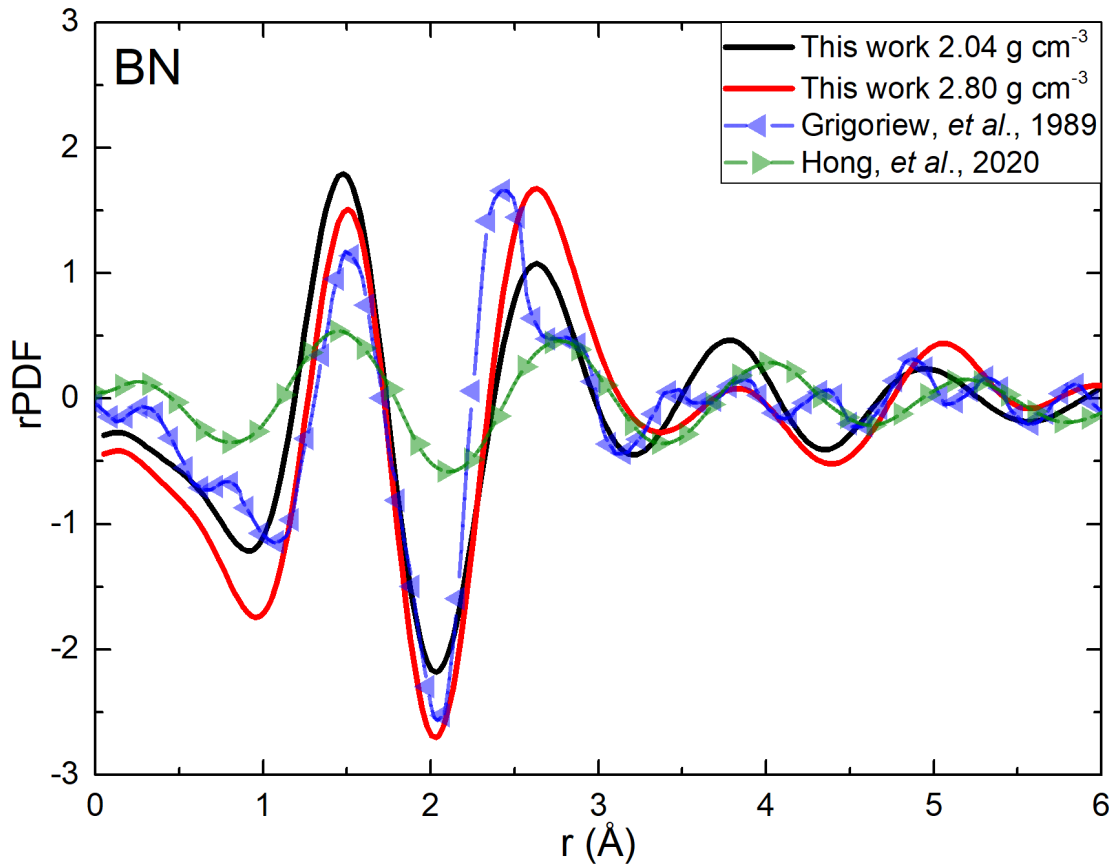

Figure 3. Total reduced Pair Distribution Functions for our two specimens and the two experimental results reported in the literature [5,22]

In 2000 Mc Culloch and collaborators applied the Car-Parrinello molecular dynamics approach to two boron-nitrogen supercells (with densities of 2.0 and 3.0 g cm$^{-3}$). To amorphize the material they liquefied it and quenched it to solid temperatures. As mentioned, doing this process may induce the appearance of some liquid characteristics in the resulting amorphous structures: Also, they used 64 atoms in the simulations as opposed to the 216 atoms used in this work. These two factors may explain the subtle differences between our results and Mc Culloch´s, presented in Figure 4, given the fact that in both works the density functional approach was applied.

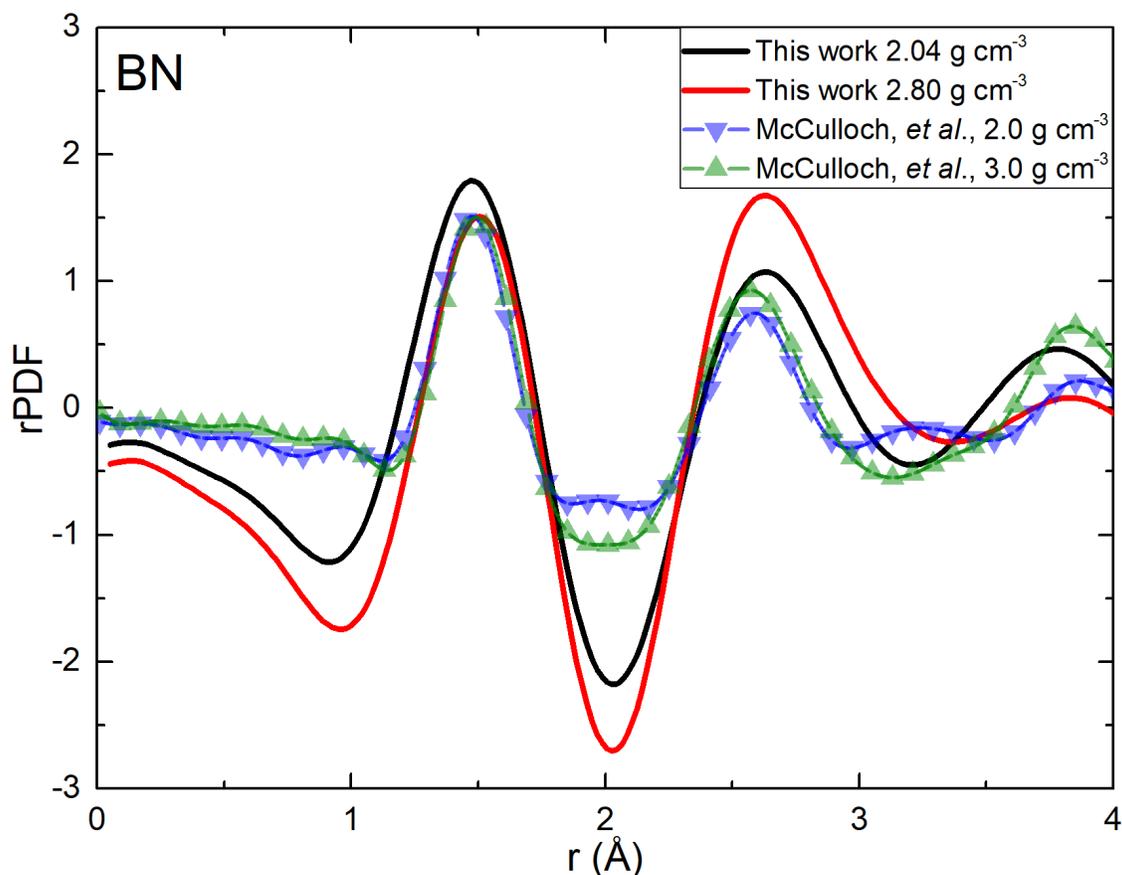

Figure 4. Comparison of the rPDFs obtained by computer simulations for samples very close in densities, 2.04 vs 2.0 and 2.80 vs 3.0, in g cm$^{-3}$ [9]

Simulational results due to Durandurdu are also included here for comparison [23,24]; the discrepancies are evident. The reason being that the ways they prepared the amorphous samples are very different from ours, and may respond to different conceptions of the amorphous solids they wanted to generate. Figures 5 for the partial PDFs make this evident. This is also the case for the PADs shown in Figures 6. Due to the fact that they reported the pPDFs and the PADs, it was possible to compare their findings with ours. The plane angle distribution functions, Figures 6, give idea of the average structures by analyzing the prominent angles that appear in the PADs.

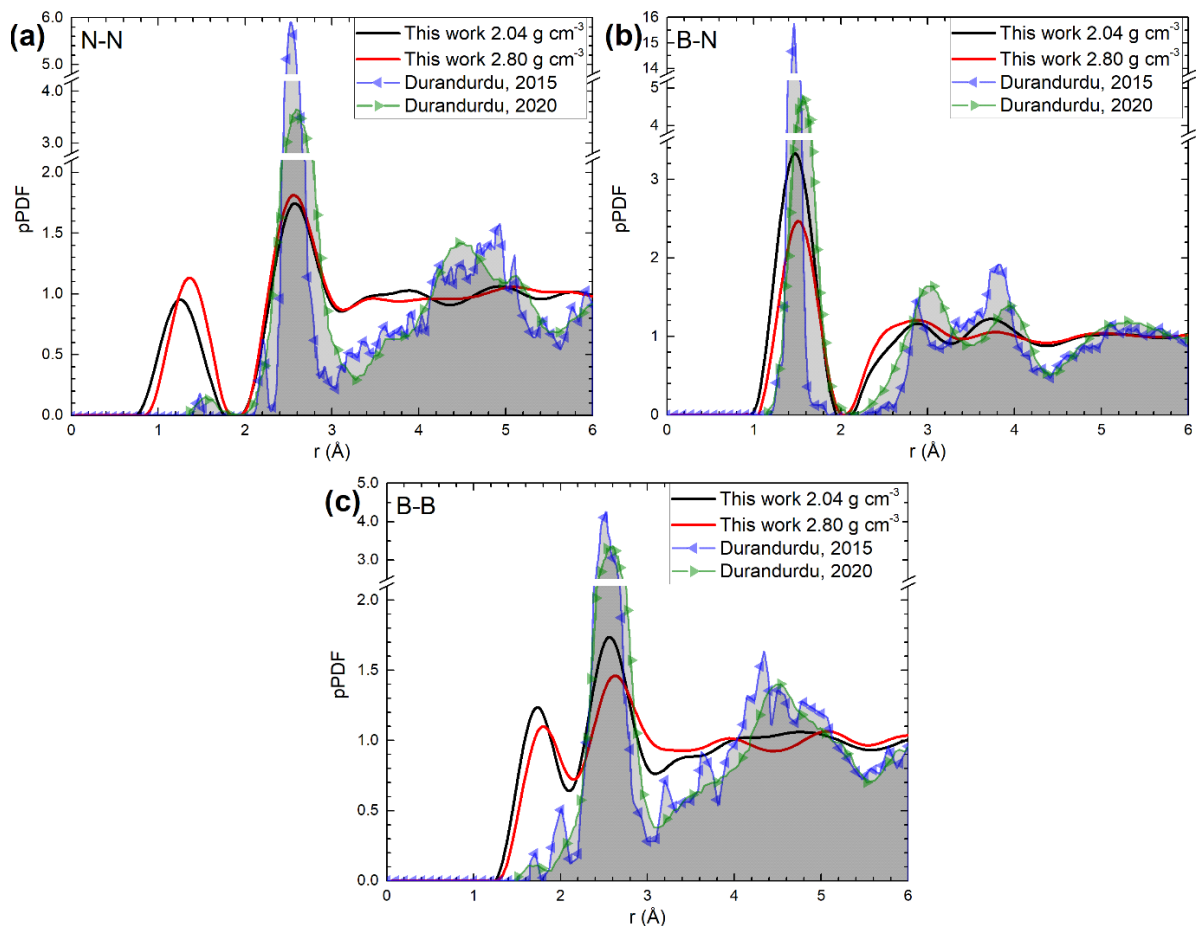

Figure 5. Comparison of Durandurdu´s results with ours for the partials a) N-N, b) B-N, and c) B-B PDFs.

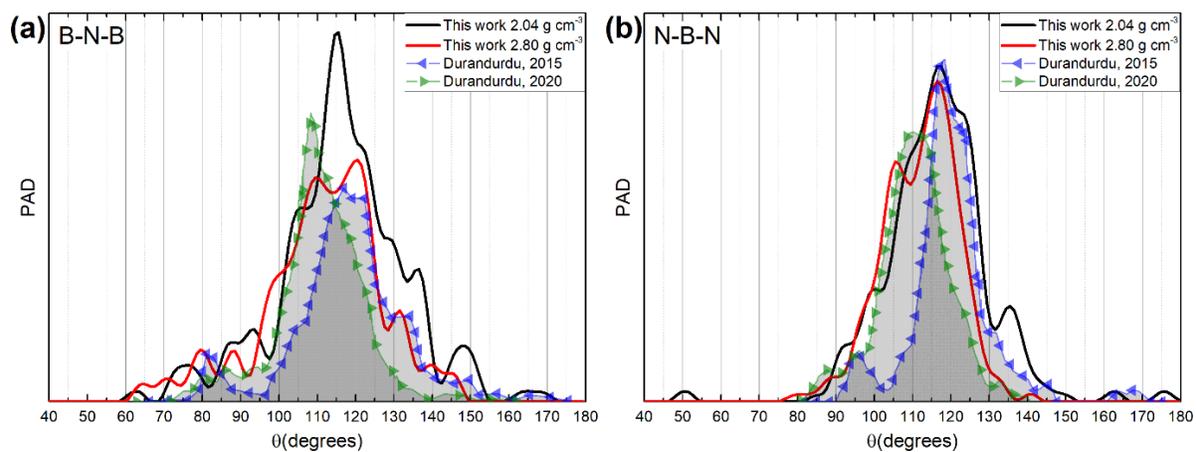

Figure 6. Comparison of Durandurdu´s results with ours for the partial PADs a) B-N-B and b) N-B-N.

### ii) The vibrational density of states

The reason for the present work was to calculate the vibrational density of states of two amorphous samples of boron-nitrogen to investigate what happens with the different crystalline vDoS during the amorphizing process. It is well established that for solids at low temperatures the Debye approximation, a non-atomistic continuous approach, may be used to describe the behavior of the vibrations. This implies that the vDoS for low $\omega$ should vary as $\omega^2$ and the specific heat at constant volume should vary as $T^3$. One would expect that if this happens for crystalline materials at low temperatures it should also happen for the "homogeneous" structure of the amorphous counterparts. However, experimentalist found that the Debye approximation does not work too well in the amorphous phases and that a high percentage of low-frequency phonon modes appear together with the decreasing and widening of the crystalline optical peaks. In 2006 these experimental results led us to study the phonon density of states in an amorphous structures of pure silicon, and the findings corroborated the arguments given above [25].

So these facts were a motivation for the study of the vDoS of the *a*-BN. By looking at Figure 7 this very same phenomenon is displayed; the vDoS of the two amorphous samples develop a larger concentration of low-frequency modes, soft-phonon modes, and the high frequency modes almost disappear. Those low-frequency modes are located around 80 meV with a timid attempt to rescue some of the high frequency modes of the material at about 140 meV; nowadays it is believed that this is a common occurrence in the transition from crystalline to amorphous. The lower density displays a small percentage of high frequency modes compared to the larger density; the same phenomenon appears for $\omega \sim 0$, which implies that the denser sample has a more localized *F(ω)*. Even more, the frequency modes above 240 meV, present for the 2.04 g cm$^{-3}$ but absent for the 2.80 g cm$^{-3}$ densities, may be a consequence of the difference in the interatomic distances, shorter for the former than the latter, for N-N and B-B (see Table 1). The modes of the amorphous samples for other densities should be investigated to discern what is causing this behavior.

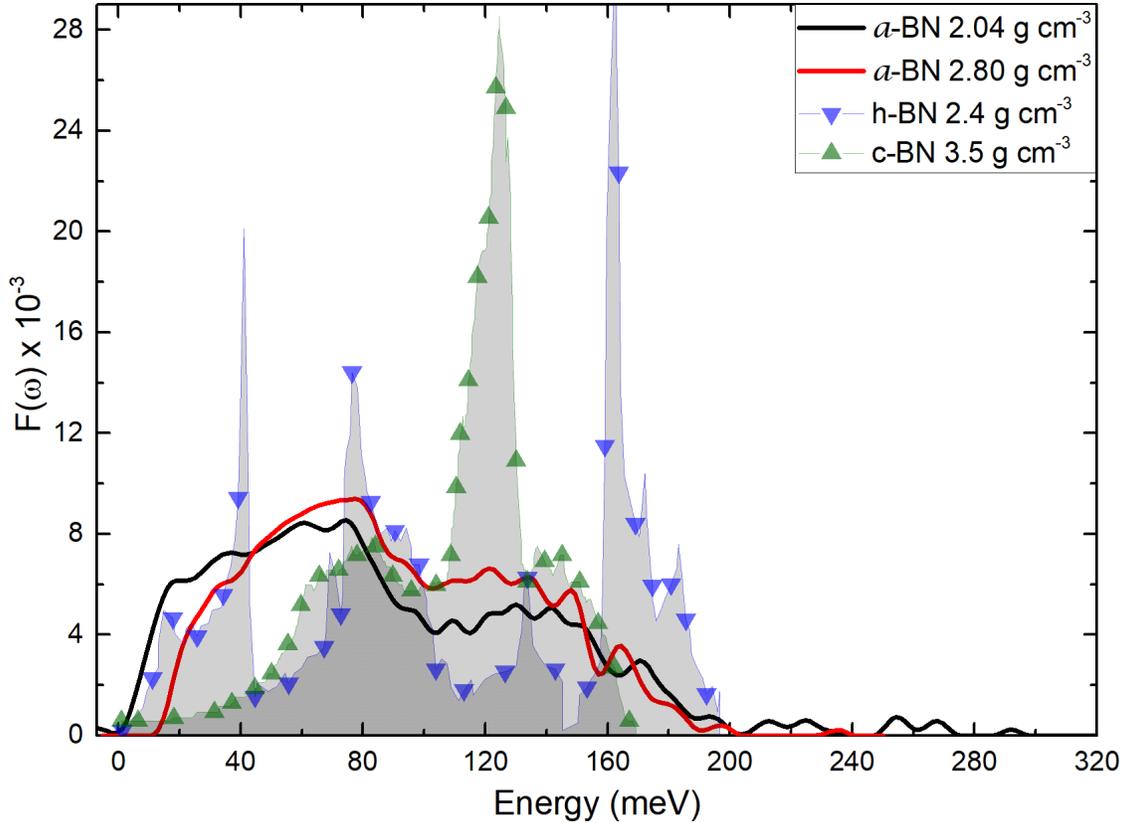

Figure 7. Vibrational densities of states for the two amorphous samples studied in this work, compared to the crystalline results for the cubic [26] and hexagonal structures [27].

### iii) Some Thermodynamics

To continue with the study of the changes that take place when a crystalline solid becomes amorphous, the internal energy and the specific heat at constant volume for our amorphous samples have been calculated. In Figures 8 the behavior for the internal energy and specific heat is observed. In Figures 9 the $C_V/T^3$ and $F(\hbar\omega)/(\hbar\omega)^2$ are plotted. At low temperatures the $C_V$ behavior looks like Debye, and at high temperatures it looks like Dulong-Petit. When the specific heat is divided by $T^3$ and plotted as a function of temperature, a straight horizontal line is expected at low temperatures, but instead a bump is observed below 10 K for the low density sample and around 60 K for the high density sample (see the insets of Figures 9). These are manifestations of the presence of the soft-phonon modes.

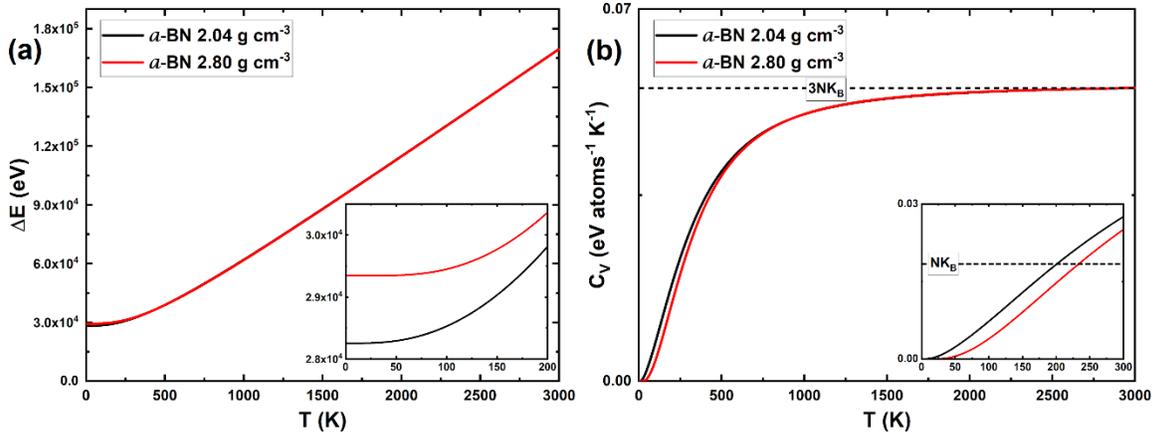

Figure 8. (a) Internal energies and (b) constant volume specific heats for the two structures studied in this work. In the insets the interval between 0 and 200 K was displayed to compare with Figure 10

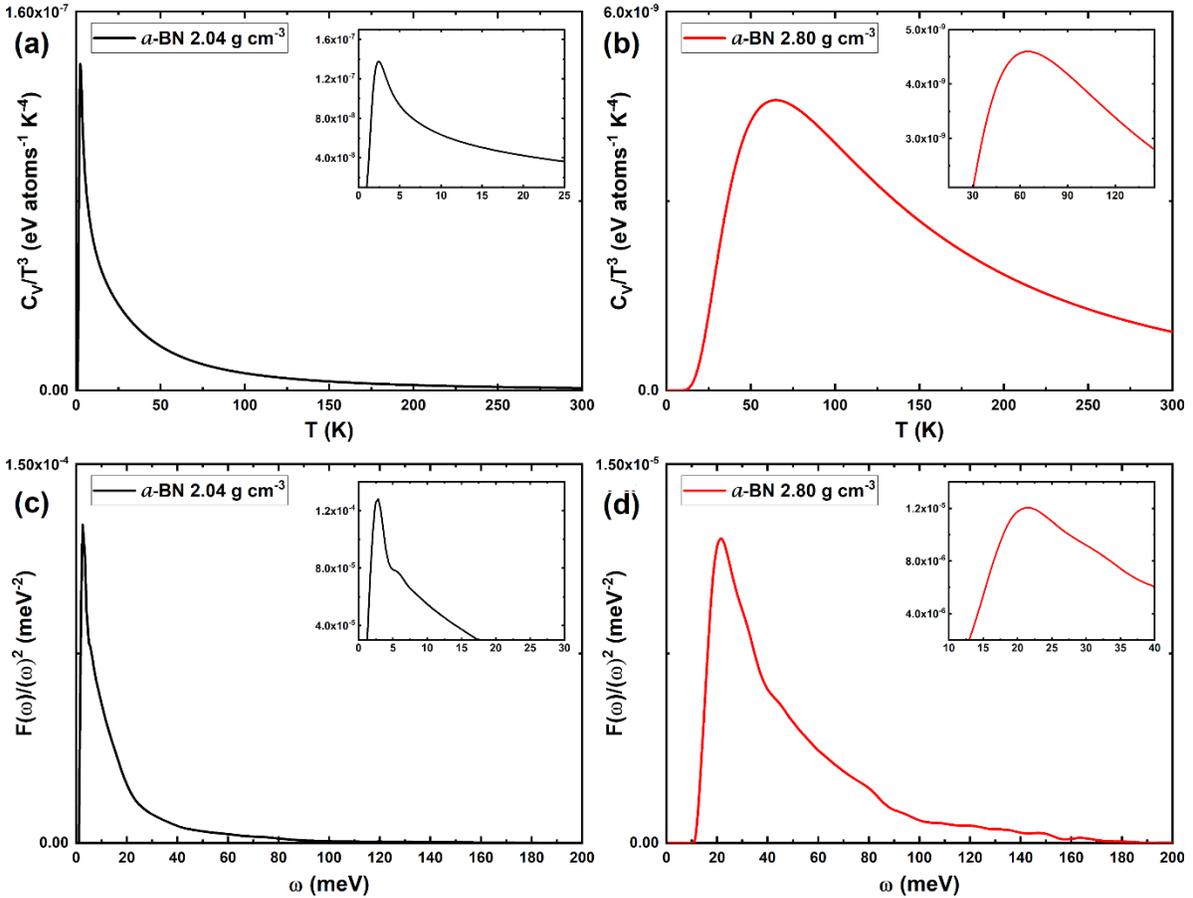

Figure 9. $C_v/T^3$ (a) for the 2.04 and (b) for the 2.80 g cm$^{-3}$ amorphous structures of this work; the insets display the temperature intervals (0, 25) K and (30, 120) K, respectively. $F(\hbar\omega)/(\hbar\omega)^2$ (c) for the 2.04 and (d) for the 2.80 g cm$^{-3}$ samples; the insets display the energy intervals (0, 30) meV and (10, 40) meV, respectively

**Conclusions**

We have studied two structures of $a$-BN, generated by MD and GO within the *undermelt-quench* procedure developed in our group. The densities investigated were 2.04 and 2.80 g cm$^{-3}$. When amorphized, most differences in the crystalline structures disappear and the topological arrangements become a "universal" feature which is modulated by the value of the density that may influence the structure and the hybridization of the chemical bond in covalent specimens. The amorphous material then is at best a function of the original density and therefore, this physical property has to be adequately chosen to describe the corresponding final amorphicity.

We calculated the correlation properties of these two samples and compared them with existing ones, both experimentally and simulationally. Our results agree with some and disagree with others. Given the limited amount of existing results for $a$-BN, it is not surprising that this happens since the statistical fluctuations in such a small universe are large. What is surprising is that there are not more investigations of these properties given the potential applications that this material is claimed to have. More studies are needed along these lines.

The vibrational properties of the structures were determined and found that the $a$-BN behaves in a similar manner to other amorphous samples; namely, the sharp features of the crystalline disappear and soft features are displayed for the amorphous. Also, soft-phonon modes play an important role in the system, while the optical modes of the crystalline fade away. The vDoS of the two samples are surprisingly similar, perhaps due to the fact that the densities considered are not very different.

Finally, the calculations of the internal energy and the constant volume specific heat clearly manifest the presence of the soft phonon modes by the appearance of a bump in the plots of $C_V/T^3$ and $F(\hbar\omega)/(\hbar\omega)^2$, typical behavior of amorphous solids.

**ACKNOWLEDGEMENTS**

David Hinojosa-Romero acknowledges Consejo Nacional de Ciencia y Tecnología (CONACyT) for supporting his graduate studies. Isaías Rodríguez thanks PAPIIT, DGAPA-UNAM, for his postdoctoral fellowship. Ariel A. Valladares, Renela M. Valladares and Alexander Valladares thank DGAPA-UNAM (PAPIIT) for continued financial support to carry out research projects under Grants No. IN104617 and IN116520. M.T. Vázquez and O. Jiménez provided the information requested. A. Lopez and A. Pompa assisted with the technical support and maintenance of the computing unit at IIM-UNAM. Simulations were partially carried at the Computing Center of DGTIC-UNAM under the project LANCAD-UNAM-DGTIC-131.